# Optically controlled single-valley exciton doublet states with tunable internal spin structures and spin magnetization generation

*Jiawei Ruan[1,2], Zhenglu Li[1,2,3], Chin Shen Ong[1,2], Steven G. Louie[1,2]\**

[1] Department of Physics, University of California at Berkeley, Berkeley, California 94720, USA

[2] Materials Sciences Division, Lawrence Berkeley National Laboratory, Berkeley, California 94720, USA.

[3] Present Address: Mork Family Department of Chemical Engineering and Materials Science, University of Southern California, Los Angeles, California 90089, USA

\*Corresponding author. Email: sglouie@berkeley.edu

**Abstract:**

Manipulating quantum states through light-matter interactions has been actively pursued in two-dimensional (2D) materials research. Significant progress has been made towards the optical control of the valley degrees of freedom in semiconducting monolayer transition-metal dichalcogenides (TMD), based on doubly degenerate excitons from their two distinct valleys in reciprocal space. Here, we introduce a novel kind of optically controllable doubly degenerate exciton states that come from a single valley, dubbed as single-valley exciton doublet (SVXD) states. They are unique in that their constituent holes originate from the same valence band, making possible the direct optical control of the spin structure of the excited constituent electrons. Combining *ab initio GW* plus Bethe-Salpeter equation (*GW*-BSE) calculations and a newly developed theoretical analysis method, we demonstrate such novel SVXD in substrate-supported monolayer bismuthene – which has been successfully grown using molecular beam epitaxy. In each of the two distinct valleys in the Brillouin zone, strong spin-orbit coupling and $C_{3v}$ symmetry lead to a pair of degenerate 1s exciton states (the SVXD states) with opposite spin configurations. Any coherent linear combinations of the SVXD in a single valley can be excited by light with a specific polarization, enabling full manipulation of their internal spin configurations. In particular, a controllable net spin magnetization can be generated through light excitation. Our findings open new routes to control quantum degrees of freedom, paving the way for applications in spintronics and quantum information science.

**Significance**

Quantum information technology relies heavily on controllable quantum states. Excitons possess essential quantum degrees of freedom, such as spins, that in principle can be optically manipulated for control and readout. The new two-dimensional topological insulator, substrate-supported monolayer bismuthene, provides an unprecedented platform to explore the internal spin structure of excitons due to its large spin-orbit coupling. Here, we discover a new type of doubly degenerate exciton states—single-valley exciton doublet states—whose constituent holes are identical. We show the internal spin structures in the coherent excitation of such excitons can be directly controlled by light polarization, thus enabling for example tunable spin magnetization generation through excitonic processes. Our work inspires new routes to quantum state control and possible applications for spintronics.

**Main Text:**

Excitons are bound electron-hole pairs[1,2]. Due to their coupling to photons, superposition of excitons can be initialized and readout by optical means, making them an attractive medium for optical manipulation of quantum degrees of freedom (such as spins) that are embedded in them[3–5]. For example, in monolayer transition-metal dichalcogenides (TMDs), two energetically degenerate excitons are located separately at the $K$ and $K'$ valleys in the Brillouin zone (BZ), carrying valley indices. They couple to left- and right-circularly polarized light, respectively[6–10]. Valley coherence can be accessed via light polarization, and its flexible manipulation has been achieved through applied magnetic field and optical approaches[11–13].

The rapidly growing family of two-dimensional (2D) materials opens up unprecedented opportunities to study new states of matter[14–17]. Recently, a monolayer of bismuth atoms supported by semiconducting silicon carbide (Bi/SiC for short) has been successfully grown using molecular beam epitaxy (MBE)[18]. Bi/SiC was reported as a 2D quantum spin Hall insulator with a large topological band gap[18–20]. The spin-orbit coupling (SOC) effect in this material is significant (with a strength of around 1 eV), largely exceeding other existing 2D materials. A bismuth monolayer on SiC (as shown in Fig. 1a) exhibits a honeycomb structure with two symmetry-related sublattices within a unit cell; meanwhile, inversion symmetry is broken due to strong hybridization with the substrate[18]. This structure is distinct from those of monolayer graphene and monolayer TMDs, the other two typical honeycomb-structured materials in the literature. Very recently, excitons were experimentally observed in high-quality Bi/SiC samples, and their resonance energies were found to be in good agreement with *ab initio GW*-BSE calculations[21]. Therefore, this newly emerged 2D material could provide an innovative platform to explore excitons with novel forms and their interplay with spins.

In this work, we show that strong SOC and $C_{3v}$ symmetry (within a valley) lead to a new type of degenerate exciton states — the single-valley exciton doublet (SVXD) states — with optically addressable internal spin configurations in Bi/SiC, making possible applications in spintronics and quantum information. Through *ab initio GW*-BSE calculations, group theoretic analysis, and pseudo-Bloch functions representations, we show that 1) the two excitons forming the SVXD arise within a *single* valley and have a 1s-like exciton envelope function; 2) their constituent holes originate from the same valence band and are spin-unpolarized, whereas their constituent electrons are distinct for the two states with opposite spin polarization and sublattice distribution; and 3) they couple to photons with opposite circular polarization of light. Notably, excitation light with a specific polarization can create a coherent superposition state of the SVXD, while controllably configuring its internal spin structure. A net spin magnetization can be generated through excitonic processes and manipulated by varying the polarization and frequency of the incident light.

Figure 1a shows the atomic structure of Bi/SiC used in our first-principles calculations of its electronic and optical properties. This interfacial structure has been successfully realized experimentally and the MBE-grown structure shows no lattice mismatch between the monolayer Bi and the SiC substrate[18]. To reduce the computational cost, we use a single layer of SiC to simulate the substrate effect in our calculations, which is shown to accurately capture the low-energy band structure and the physics in this paper (See SI for more details). Figure 1b depicts the Kohn-Sham band structure computed using density functional theory (DFT) in the local density approximation (LDA) as well as the quasiparticle band structure computed using the *ab initio GW* method[22]. Both calculations include SOC within the full-spinor formalism. At the $K$ ($K'$) point, the direct gap is 0.93 eV at the DFT level and is 1.64 eV at the *GW* level. There is a large energy splitting between the topmost two valence bands (of 0.43 eV at the *GW* level). In contrast, the bottom-most two conduction bands at the $K$ ($K'$) point are degenerate (see detailed discussion below). The wavefunctions of the states of these four (two conduction and two valence) low-energy bands at $K$ and $K'$ valleys (the two valleys are related by time-reversal symmetry) are basically localized in the bismuth monolayer, with a major contribution from the bismuth $p_x$ and $p_y$ orbitals and a minor contribution from the bismuth $s$ orbital[18].

The significant SOC and substrate effects in this system induce intriguing spin-orbit entangled properties to the low-energy electronic states at the *K* and *K'* valleys[19,23,24]. In the absence of SOC, a graphene-like Dirac gapless dispersion (instead of a gap in the presence of SOC) appears at $K$ (and $K'$)[19,23]. At the Dirac energy point in the case of no SOC, the states at $K$ are composed of orbitals that are $p_+^A$ and $p_-^B$ associated with $A$ and $B$ sublattices of Bi atoms in the honeycomb structure ($p_\pm = (\mp p_x - ip_y)/\sqrt{2}$). Combining the local on-site orbital angular momentum ($\pm 1$ for $p_\pm$) and lattice angular momentum ($\pm 1$ for A/B sublattice at wavevector **K**) together, the $p_+^A$ and $p_-^B$ states carry total orbital angular momenta (in units of $\hbar$) $L_z = \pm 2 \equiv \mp 1 (\text{mod } 3)$, respectively[23]. Turning on SOC, the electron spins of these basis orbitals mix in and a low-energy Hilbert space of dimension-4 can be constructed. Using the basis of $(|p_+^A \uparrow, \mathbf{K}\rangle, |p_+^A \downarrow, \mathbf{K}\rangle, |p_-^B \uparrow, \mathbf{K}\rangle, |p_-^B \downarrow, \mathbf{K}\rangle)$ ($\uparrow/\downarrow$ denotes spin up/down) with total angular momenta $J_z = 5/2, 3/2, -3/2, -5/2$, respectively, the effective low-energy Hamiltonian at $K$ reads[23]

$$H^{\text{SOC}}(\mathbf{K}) = \begin{pmatrix} \lambda_{\text{SOC}} & 0 & 0 & 0 \\ 0 & -\lambda_{\text{SOC}} & -i\lambda_R & 0 \\ 0 & i\lambda_R & -\lambda_{\text{SOC}} & 0 \\ 0 & 0 & 0 & \lambda_{\text{SOC}} \end{pmatrix}, \quad (1)$$

where $\lambda_{\text{SOC}}$ is an on-site SOC matrix element and is positive, making $|p_+^A \uparrow, \mathbf{K}\rangle$ and $|p_-^B \downarrow, \mathbf{K}\rangle$ to be the conduction band states with the same energy, as well as making the linear combinations of $|p_+^A \downarrow, \mathbf{K}\rangle$ and $|p_-^B \uparrow, \mathbf{K}\rangle$ to be valence band states. $\lambda_R$ (a real number) originates from a Rashba-type SOC, induced by an out-of-plane effective electric field generated from the SiC substrate, and mixes $|p_+^A \downarrow, \mathbf{K}\rangle$ with $|p_-^B \uparrow, \mathbf{K}\rangle$. Thus, the two valence eigenstates at **K** take the form of $(|p_+^A \downarrow, \mathbf{K}\rangle \pm i|p_-^B \uparrow, \mathbf{K}\rangle)/\sqrt{2}$. We emphasize that the two topmost valence states are split in energy, whereas the bottom two conduction states remain degenerate, which is guaranteed by the $C_{3v}$ symmetry at the $K$ point. This band arrangement hosts unique exciton structures as will be discussed below. Using time-reversal (TR) symmetry, the band states at $K'$ can be obtained, with $|p_-^A \downarrow, \mathbf{K}'\rangle$ and $|p_+^B \uparrow, \mathbf{K}'\rangle$ as the two degenerate conduction band states and $(|p_+^B \downarrow, \mathbf{K}'\rangle \mp i|p_-^A \uparrow, \mathbf{K}'\rangle)/\sqrt{2}$ as the two split valence band states, which is equivalent to switching $A$ and $B$ sublattice notations of the states at $K$. Away from the $K/K'$ points, a **k**-dependent Rashba SOC effect[23] leads to a small but anisotropic Rashba-type splitting in the energy of the two conduction bands (Fig. 1c ). This splitting is much smaller than the band gap, but it leads to a spin character mixing of the conduction band states near the $K/K'$ points. Here we emphasize that, the novel single-valley exciton doublet states we discuss below is formed by the degenerate states within a single individual valley. The photoresponse of excitons in the *K* and *K'* valleys (TR-related) are identical, and the whole system follows the single-valley physics.

By directly diagonalizing the first-principles *GW*-BSE Hamiltonian[25,26], a series of exciton states (their energies and wavefunctions) are obtained with detailed information on their wavefunction character. The *ab initio* computed optical absorption spectrum for linearly polarized light for our system, including electron-hole interaction (excitonic) effects, is displayed in Fig. 1d (red solid curve) as compared to the non-interacting-particle result (blue dashed curve). There are two main peaks, denoted as A and B, with excitation energies of 1.22 eV and 1.64 eV, respectively. Peak A (B) corresponds to the excitation of the lowest-energy optically bright exciton states that are mainly formed by interband transitions between the first (second) valence band and the two degenerate conduction bands. The excitons from both valleys (related by TR symmetry from each other) contribute to the peaks for this case of linearly polarized light.

We now investigate in detail the character of the SVXD in one specific valley, which is allowed because the coupling between $K$ and $K'$ valleys is negligible (~0.1 meV, see Supplementary Information (SI)). We focus on the A-series exciton properties in the $K$ valley. Fig. 2a shows the energies and optical selection rules of the exciton states for linearly polarized light. To characterize their symmetry properties and degeneracies, we developed a method to directly calculate the group representations for the exciton states (See SI for the details). The irreducible representations under $C_{3v}$ symmetry for several of the low-energy excitons are shown in Fig. 2a.

The lowest two excitons $|1s_+\rangle$ and $|1s_-\rangle$ of the K valley are degenerate, corresponding to $\mathbf{e}_+$ and $\mathbf{e}_-$ chirality, respectively, where $\mathbf{e}_\pm$ are defined as $(\mp \mathbf{e}_x - i\mathbf{e}_y)/\sqrt{2}$ with $\mathbf{e}_x$ ($\mathbf{e}_y$) denoting the unit vector along the $+x$ ($+y$) direction, defined in a global coordinate system (see Fig. 1a). The nature of these excitons, however, can be obscured by the direct numerical diagonalization from the $GW$-BSE calculations, which contains challenges associated with the random phases and mixing of degenerate states (see SI). To solve this problem, we adopt the concept of pseudo-Bloch functions[27] (denoted as $|\chi_{m,\mathbf{k}}\rangle$ for the single-particle orbitals, which is a symmetrized combination of Bloch states at a given $\mathbf{k}$ point as shown in Fig. 2b). They serve as a new basis of electron and hole states with a smooth gauge to give unambiguous representations of the internal structure of exciton states. The constructed two pseudo-Bloch conduction band states $|\chi_{c1,\mathbf{k}}\rangle$ and $|\chi_{c2,\mathbf{k}}\rangle$ have main character of $p_+^A \uparrow$ (denoted as $\Uparrow$) and $i p_-^B \downarrow$ (denoted as $\Downarrow$), respectively. These pseudo-Bloch conduction band states have spins that are locked with sublattice indices and angular momentums, which inherit the properties of the states at the $K$ point. The energy bands depicted in Fig. 2b for the pseudo-Bloch states are defined as the expectation value of these constructed states with respect to the quasiparticle Hamiltonian.

In the new pseudo-Bloch basis, the degenerate exciton states of the A series can be expressed as $|1s_\pm\rangle = \sum_\mathbf{k} C_1^{1s_\pm}(\mathbf{k})|\chi_{v\mathbf{k}}^*; \chi_{c1,\mathbf{k}}\rangle + \sum_\mathbf{k} C_2^{1s_\pm}(\mathbf{k})|\chi_{v\mathbf{k}}^*; \chi_{c2,\mathbf{k}}\rangle$, where $C_i^S(\mathbf{k})$ are the envelope functions in this constructed new basis (see SI for details), as shown in Fig. 2d. We find that the $1s_+$ and $1s_-$ excitons are dominated (>99.7%) by transitions from the topmost valence band to the first and second pseudo-Bloch conduction bands shown as the upper and lower panels in Fig. 2c, respectively, with a same $s$-like envelope function in $\mathbf{k}$-space. Thus, they can be simply written as $|1s_+\rangle \approx \sum_\mathbf{k} f(\mathbf{k}) |\chi_{v\mathbf{k}}^*; \chi_{c1,\mathbf{k}}\rangle$ and $|1s_-\rangle \approx \sum_\mathbf{k} f(\mathbf{k}) |\chi_{v\mathbf{k}}^*; \chi_{c2,\mathbf{k}}\rangle$, where $f(\mathbf{k}) \approx C_1^{1s_+}(\mathbf{k}) \approx C_2^{1s_-}(\mathbf{k})$.

The optical selection rules[28,29] for the SVXD become clear in this new basis. Taking the $1s_+$ exciton as an example, the interband velocity matrix elements for its dominant interband transition ($v \to c1$) are related to $\tilde{v}^{\mp}_{v,c1;\mathbf{k}} = \langle \chi_{v,\mathbf{k}} | \mathbf{e}_{\mp} \cdot \hat{\mathbf{v}} | \chi_{c1,\mathbf{k}} \rangle = \langle \chi_{c1,\mathbf{k}} | \mathbf{e}_{\pm} \cdot \hat{\mathbf{v}} | \chi_{v,\mathbf{k}} \rangle^* = (\tilde{v}^{\pm}_{c1,v;\mathbf{k}})^*$, which correspond to the optical excitation from the top valence band to the first pseudo conduction band illuminated by $\mathbf{e}_{\pm}$ circularly polarized light, respectively. These matrix elements viewed as a 2D vector field are shown in the upper two panels in Fig. 2e for the two circular polarizations. They have winding numbers $l_- = 0$ and $l_+ = 2$, respectively (see SI for more details about choosing the conventions in the winding numbers and the selection rules). Comparing these winding numbers to the angular momentum of the envelope function of the $1s_+$ exciton being $m = 0$, we see for this exciton we have $m = -l_-$ and $m \neq -l_+$. A generalized optical selection rule for excitons in 2D systems is recently derived by Cao, Wu and Louie (CWL)[28] which says that an exciton is optically dark unless $m = -l_{\mp}$ (mod $n$) for a material with $n$-fold rotational symmetry along with sufficient magnitude for velocity matrix elements. Following the CWL selection rule, we can conclude that $|1s_+\rangle$ is strongly optically active under $\mathbf{e}_+$ polarized light and strictly optically inactive under $\mathbf{e}_-$ polarized light. Similarly, the CWL selection rule together with the results in Fig. 2e shows that the $1s_-$ exciton is strongly coupled to $\mathbf{e}_-$ polarized light only.

Here, we propose that the singe-valley exciton doublet states and their internal spin configurations can be coherently controlled by optical means because of their optical selection rules and special internal structures noted above. We first introduce the Bloch sphere to describe light with a general in-plane polarization such as elliptical polarization (Fig. 3a). Given the two-angle parameter $\boldsymbol{\vartheta} = (\theta, \phi)$, the corresponding unit vector of the light polarization is defined as $\mathbf{e}_{\boldsymbol{\vartheta}} = \cos\frac{\theta}{2} \mathbf{e}_+ + \sin\frac{\theta}{2} e^{i\phi} \mathbf{e}_-$. Under illumination of such polarized light, a coherent superposition of the SVXD can be excited in the form of $|1s_{\boldsymbol{\vartheta}}\rangle = \cos\frac{\theta}{2} |1s_+\rangle + \sin\frac{\theta}{2} e^{i\phi} |1s_-\rangle$. Since the basis exciton states $|1s_+\rangle$ and $|1s_-\rangle$ basically share the *same* envelope function and *same* hole states (in the pseudo-Bloch basis), this coherent exciton state can be further expressed as

$$|1s_{\boldsymbol{\vartheta}}\rangle \approx \sum_{\mathbf{k}} f(\mathbf{k}) |\chi^*_{v,\mathbf{k}}\rangle \otimes |\chi_{c,\mathbf{k}}(\boldsymbol{\vartheta})\rangle \qquad (2)$$

with $|\chi_{c,\mathbf{k}}(\boldsymbol{\vartheta})\rangle = \cos\frac{\theta}{2} |\chi_{c1,\mathbf{k}}\rangle + \sin\frac{\theta}{2} e^{i\phi} |\chi_{c2,\mathbf{k}}\rangle$. We can see in this superposition state of the SVXD, the hole part $|\chi^*_{v,\mathbf{k}}\rangle$ is fixed and independent of the polarization of the incoming photon, whereas the electron part $|\chi_{c,\mathbf{k}}(\boldsymbol{\vartheta})\rangle$ is a coherent superposition of two spin-polarized electron states (right panel in Fig. 3b), which is controllable by the incident light polarization (see the schematic in Fig. 3f). In other words, the absorbed light information is directly

transferred to the excited exciton through its electron component. As a specific illustration, let us take the superposition parameters to be along a fixed longitude of the Bloch sphere in Fig. 3a (with $\phi = 0°$ as an example). The one-to-one correspondence between the $\mathbf{e}_\vartheta$ polarized light and the superposition state $1s_\vartheta$ leads to the prediction that the square modulus of the exciton velocity matrix element $T_\vartheta = |\langle 1s_\vartheta | \mathbf{e}_\vartheta \cdot \hat{\mathbf{v}} | 0 \rangle|^2$ is identical to that of $|1s_+\rangle$ under $\mathbf{e}_+$ polarized light (denoted as $T_0$) for all parameter choice $\vartheta$, shown as the red line in Fig. 3d. At the bottom of Fig. 3d, we depict the characters of internal electron (up to a $U(1)$ phase) of four special superposition states that are respectively excitable by light with $\mathbf{e}_+, \mathbf{e}_x, \mathbf{e}_-$ and $\mathbf{e}_y$ polarizations.

We evaluate the z-direction total spin expectation values $s_z^{\text{tot}}$ for the superposition states of the SVXD. For the exciton state $|1s_+\rangle$ (with the character $\frac{1}{\sqrt{2}}(|p_+^A \downarrow\rangle + i|p_-^B \uparrow\rangle) \otimes |p_+^A \uparrow\rangle$) excited by $\mathbf{e}_+$ polarized light, the hole part has zero net spin expectation, and thus $s_z^{\text{tot}}$ is solely determined by the electron part, which is around $\hbar/2$. Indeed, our first-principles $GW$-BSE calculations give $s_z^{\text{tot}}(1s_+) = 0.99 \, (\hbar/2)$. This is entirely distinct from monolayer TMDs, where there is no net spin expectation value for exciton created under circularly polarized light. For a general superposition state $|1s_\vartheta\rangle$ with superposition parameter $\vartheta = (\theta, \phi)$, the spin expectation value is given by $s_z^{\text{tot}}(1s_\vartheta) = 0.99 \cos\theta \, (\hbar/2)$, depending only on the polar angle $\theta$. We show this spin property of the superposition states along the above-mentioned excitation path in Fig. 3d.

The SVXD at the $K'$ valley can be controlled in the same manner as those at the $K$ valley (Fig. 3c). We note that in this valley, the upward spin in the excited electrons is also locked with positive orbital angular momentum (but locked with the $B$ sublattice), and it is also excited by $\mathbf{e}_+$ polarized light; and vice versa for the downward spin. Thus, the spin expectation values of the superposition states of SVXD at the $K'$ valley are the same as those at the $K$ valley for a given light polarization (see Figs. 3e and 3d). When optically excited, the two valleys have additive contributions to the spin polarization.

With this knowledge of manipulating the spin configurations of the SVXD, we propose a new optical spin generation mechanism and quantitatively estimate the amount of the generated spin magnetization. Let us consider the exciton spin expectation value in the optical absorption process. We define a spin-resolved optical response function, which measures spin magnetic moment created per unit time under illumination of unit power of light. It is given as,

$$\eta_\vartheta^z(\omega) = C_\mu \sum_S \sum_{S'} \frac{1}{\omega^2} \langle 0 | \mathbf{e}_\vartheta^* \cdot \hat{\mathbf{v}} | S \rangle \langle S' | \mathbf{e}_\vartheta \cdot \hat{\mathbf{v}} | 0 \rangle \langle S | s_z^{\text{tot}} | S' \rangle D_{S,S'} \delta(\Omega_S - \hbar\omega), \quad (3)$$

where $\hat{s}_z^{\text{tot}}$ is the z-component of the total spin operator including contributions from the electron and hole in an exciton, $\Omega_S$ is the excitation energy of the exciton states $|S\rangle$, $C_\mu$ is a constant prefactor and $D_{S,S'}$ is a function which equals 1 when $S$ and $S'$ are degenerate otherwise vanishes (see SI for more details about the derivation of Eq. (3) and the constant prefactor $C_\mu$). In Fig. 3g, we plot $\eta_{\hat{\vartheta}}^z(\omega)$ as a function of the polar angle $\theta$, with $\phi$ fixed to zero, without loss of generality. A peak appears at the resonance frequency of the A excitons, and the rate of spin magnetization creation is maximal under circularly polarized light and can be continuously tuned by varying the light polarization. Similar behaviors can be found at the resonance frequencies of the B excitons. These characteristic peaks arise because electron-hole interactions give rise to constructive interference of various spin matrix elements, resulting from excitonic wavefunctions being correlated superposition of free electron-hole pair excitations. This effect strongly enhances the light-induced spin magnetization, akin to the excitonic enhancement in the linear optical absorption shown in the Fig. 1d. We expect that this controllable spin magnetization can be detected in various photocurrent measurements[30–33].

In summary, through *ab initio GW*-BSE calculations and a newly developed pseudo-Bloch function representation method, we have predicted the existence of unique single-valley exciton doublet states whose internal spin configuration can be optically controlled in the 2D topological material Bi/SiC. The tunable spin configuration permits the optical generation and manipulation of net spin magnetization. As they offer an exceptional interface between photons and spins, we anticipate that these novel degenerate exciton states can pave the way for potential applications in quantum information science and spintronics.

**Acknowledgments:** This work was supported by the Theory of Materials Program at the Lawrence Berkeley National Laboratory (LBNL) through the Office of Basic Energy Sciences, U.S. Department of Energy under Contract No. DE-AC02-05CH11231, which provided the theoretical formulation and the GW and GW-BSE calculations. Advanced code developments were provided by the Center for Computational Study of Excited-State Phenomena in Energy Materials (C2SEPEM) at LBNL, which is funded by the U.S. Department of Energy, Office of Science, Basis Energy Sciences, Materials Sciences and Engineering Division under Contract No. DE-AC02-05CH11231, as part of the Computational Materials Sciences Program. Computational resources were provided by National Energy Research Scientific Computing Center (NERSC), which is supported by the Office of Science of the US Department of Energy under contract no. DE-AC02-05CH11231, Stampede2 at the Texas Advanced Computing Center (TACC), The University of Texas at Austin through Extreme Science and Engineering Discovery Environment (XSEDE), which is supported by National Science Foundation under grant no. ACI-1053575 and Frontera at TACC, which is supported by the National Science Foundation under grant no. OAC1818253. The authors thank Q. Li, Z. Zhang, M. Wu and T. Cao for helpful discussions.

**Author contributions:** S.G.L. and J. R. conceived the original idea for the project and S.G.L directed the research. J.R. performed the calculations and developed methods/codes for the analysis of exciton states. All authors analyzed the data, discussed the results, and wrote the paper.

**Competing interests:** The authors declare no competing interests.

**Data and materials availability:** The data that support the findings of this study are available from the corresponding author upon reasonable request.

**Corresponding author:** Correspondence and requests for materials should be addressed to sglouie@berkeley.edu

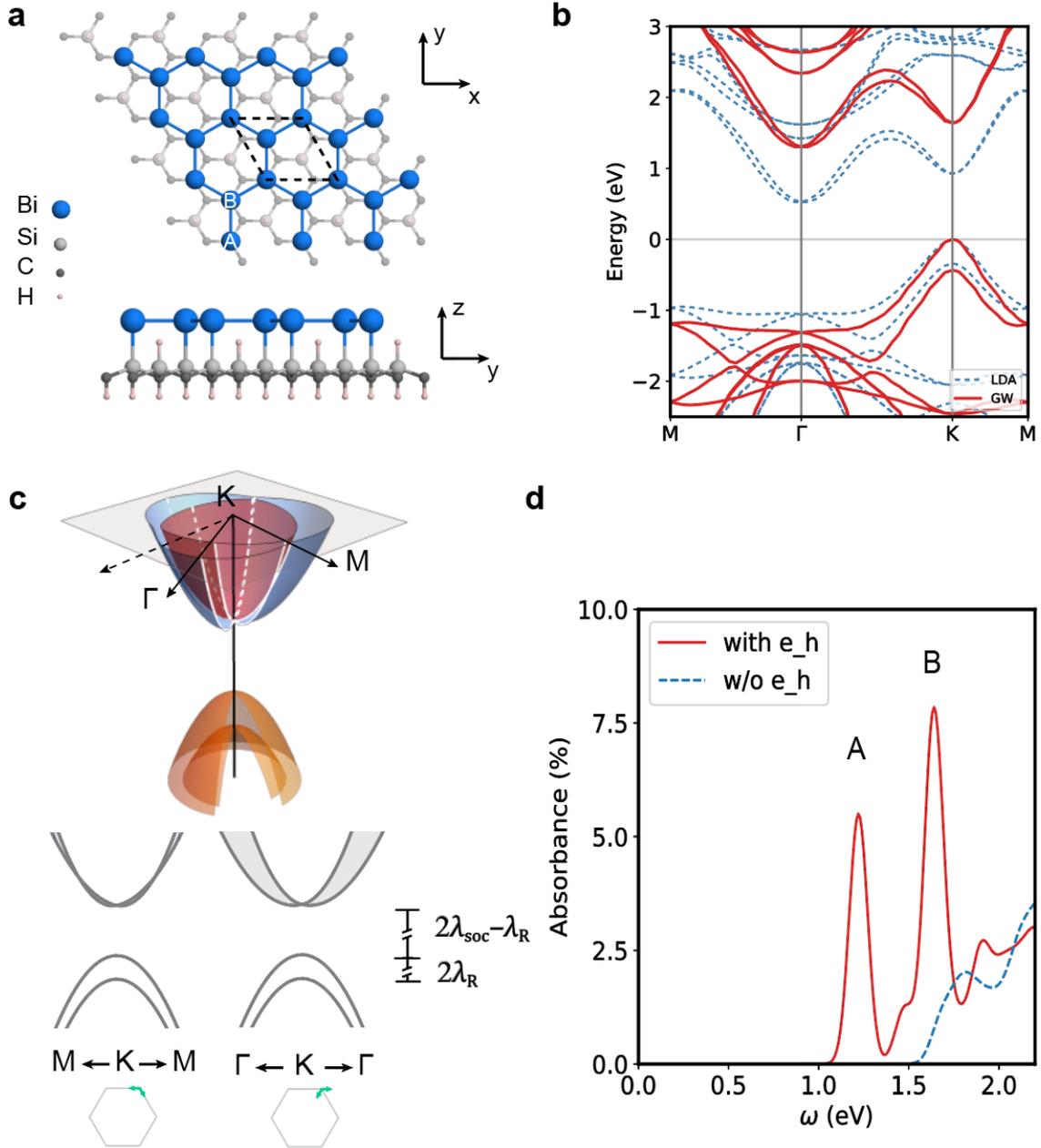

**Fig. 1. Crystal structure, electronic structure and optical absorption spectrum of substrate-supported bismuthene. a**, Top and side views of the schematic structure of monolayer bismuthene (blue) on the SiC(0001) substrate (gray). The unit cell is shown by dashed line segments. *A* and *B* sublattices of Bi atoms are indicated. **b**, Computed DFT-LDA (dashed blue) and *GW* (solid red) band structures. **c**, Schematic low-energy band structure at the *K* valley. The Rashba splitting of the two conduction bands is anisotropic (the gray filled area). The two conduction bands cross each other along the *K-M* direction and remain splitting along the *K-Γ* direction. The **k**-paths for the schematic band structures are denoted by the green arrows. **d**, Calculated optical absorption spectrum with (red) and without

(dashed blue) electron-hole interactions. A Gaussian broadening factor $\sigma = 50$ meV is used. The two main exciton peaks are labeled as A and B.

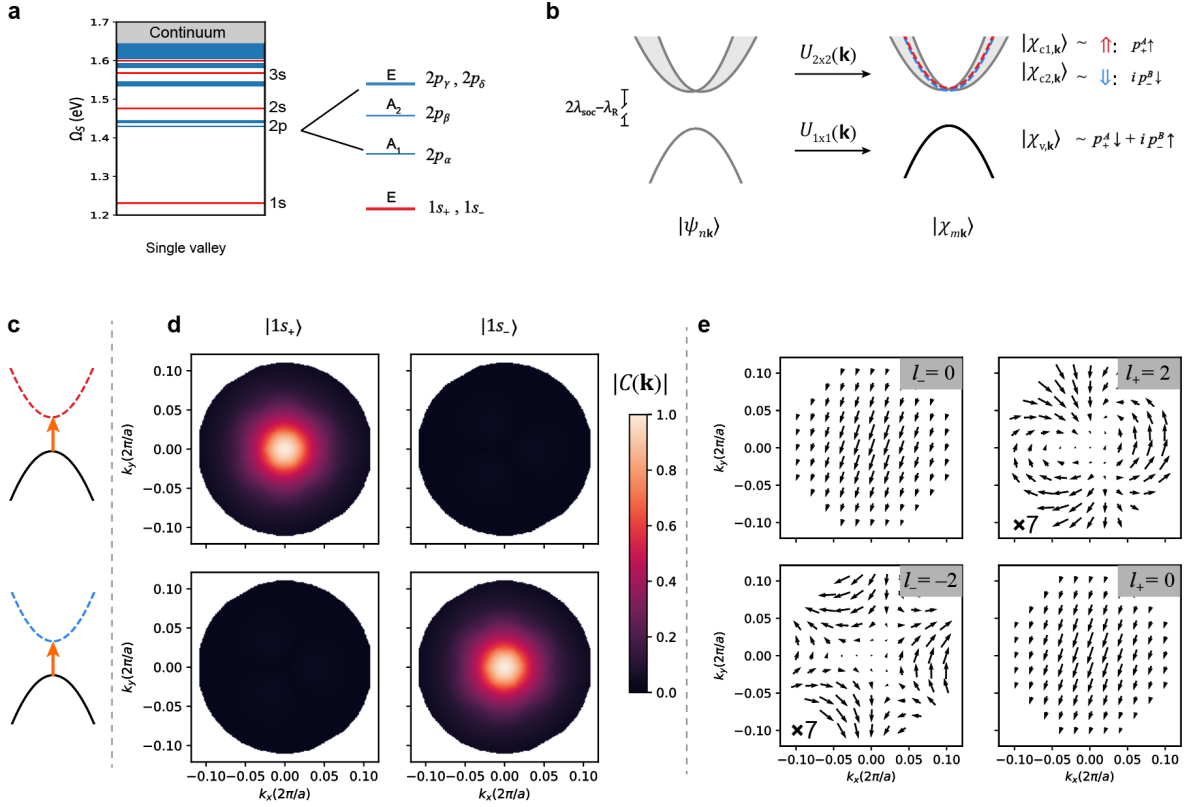

**Fig. 2. Exciton envelope functions and interband velocity matrix elements in the pseudo-Bloch basis. a**, Single-valley (the K valley as an example) exciton energy levels in the A series. Optically bright (dark) exciton states under linearly polarized light are in red (blue) color. The group representations of the lowest six exciton states are labeled in the right panel. **b**, Pseudo-Bloch functions $|\chi_{m\mathbf{k}}\rangle$ with smooth gauges, constructed from Bloch eigenfunctions $|\psi_{n\mathbf{k}}\rangle$ of bands that are degenerate at $K$ by unitary transformations. The constructed pseudo-Bloch conduction bands are depicted using dashed red and dashed blue curves, with the colors respectively denoting spin-up and spin-down polarizations. The constructed topmost valence band is depicted using a solid black curve. Spin-orbit-sublattice locked characters are shown for each of the pseudo-Bloch bands. For details of the construction method see SI Sec. S3. **c**, Illustration of the two sets of interband transitions in the pseudo-Bloch basis forming the A excitons. **d**, The absolute value of the amplitude of the envelope functions $C_i^S(\mathbf{k})$ of the excitons $1s_+$ (left top/bottom panels) and $1s_-$ (right top/bottom panels) in terms of the two pseudo-Bloch band-to-band transitions shown in **c** (top/bottom panels). **e**, The interband velocity matrix elements $\tilde{v}_{v,c;\mathbf{k}}^{\mp}$ in the basis of pseudo-

Bloch functions and their winding numbers $l_\mp$. Upper (lower) two panels are velocity matrix elements corresponding to interband transitions shown in the upper (lower) panel in **c**. The direction and length of the arrows denote the phase and magnitude of the matrix elements, respectively. The magnitudes in the lower-left and upper-right panels are multiplied by a factor of 7 for better visibility. For all the **k**-space plots in **d** and **e**, the $K$ point is placed at the origin.

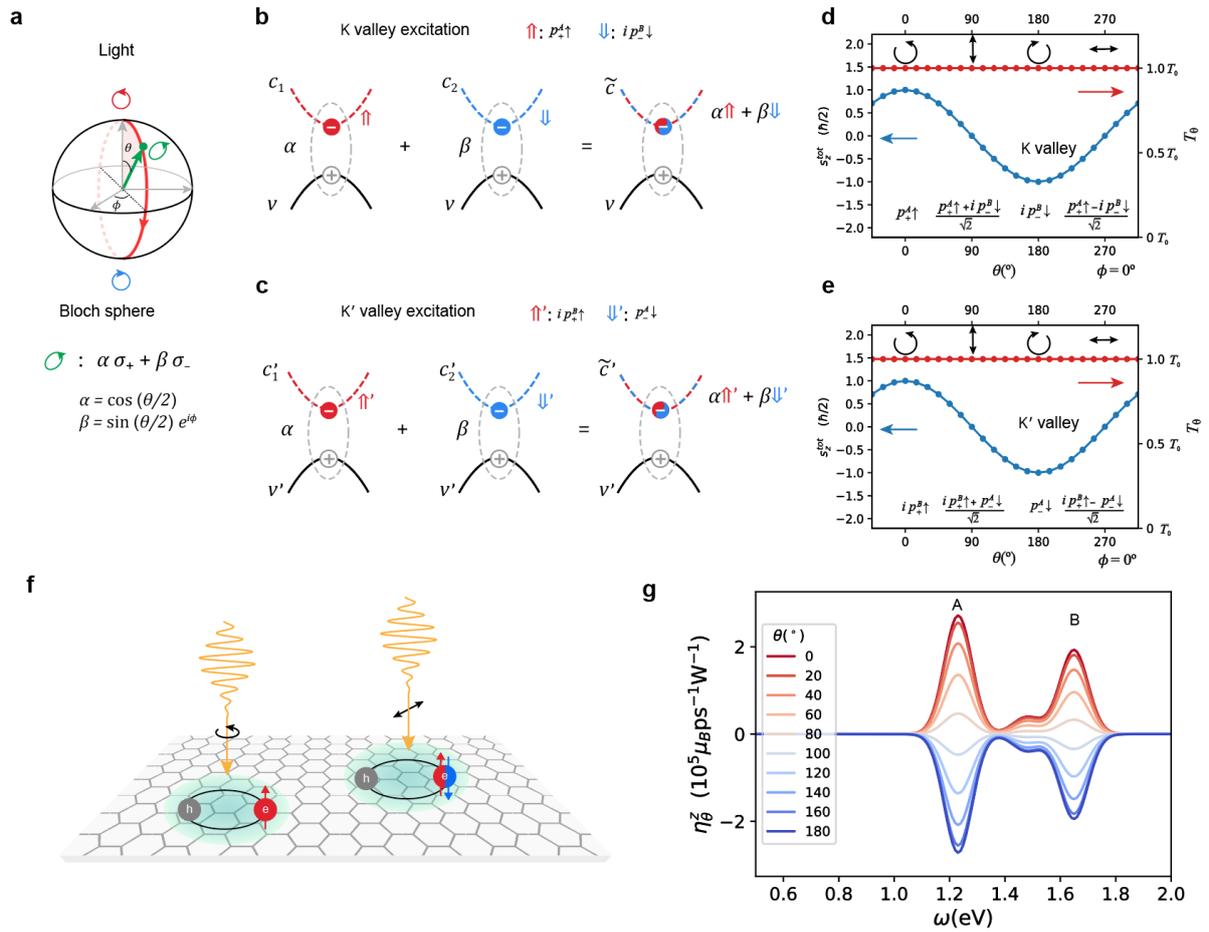

**Fig. 3. Optically excited coherent superposition of single-valley exciton doublet (SVXD) states and light controllable generation of spin magnetization. a,** General polarization of light characterized by the two-angle parameter $\vartheta = (\theta, \phi)$, which can be represented by a point (green dot) on the Bloch sphere. **b,** a coherent superposition of the SVXD states at the $K$ valley created by light with the polarization shown in **a**. This coherent exciton state can be equivalently represented as the hole bound to a coherent superposition of two electron states with the same coherent parameter. The spin-orbit-sublattice locked characters (denoted as ⇑ and ⇓) are shown at the top. **c**, same as **b** but for states at the $K'$

valley. **d**, **e**, Square modulus of the exciton velocity matrix elements $T_\vartheta$ (red) and the spin expectation values $s_z^{tot}$ (blue) of the coherent superposition exciton states at the $K$ valley and $K'$ valley, respectively. The superposition parameters are chosen to be along the longitude line of the Bloch sphere shown in **a** with $\phi = 0$. Four specific in-plane light polarizations ($\mathbf{e}_+, \mathbf{e}_y, \mathbf{e}_-$ and $\mathbf{e}_x$ polarizations) and their corresponding excited electron characters (each up to a phase) are depicted. **f**, Schematic of excitons (and their spin structures) excited by two representative light polarizations, $\mathbf{e}_+$ circularly polarized light (left panel) and linearly polarized light (right panel). **g**, Light-generated spin magnetic moment per picosecond per unit of incident light power $\eta_\vartheta^z(\omega)$ as a function of the polarization and frequency of light. A Gaussian broadening factor $\sigma = 50\,\text{meV}$ is used.